\documentclass[twocolumn]{aastex701}
\usepackage{tabularx}
\usepackage{moresize}
\usepackage{amsmath}
\usepackage{hyperref}

\begin{document}

\title{Transit Timing of the White Dwarf-cold Jupiter System WD\,1856+534}

\author[orcid=0009-0002-8671-4858, gname='Eli',sname='Gendreau-Distler']{Eli A. Gendreau-Distler}
\affiliation{Department of Astronomy, University of California, Berkeley, CA 94720-3411, USA}
\email[show]{egendreaudistler@berkeley.edu}  

\author[orcid=0009-0002-8376-7882, gname='Kate', sname='Bostow']{Kate B. Bostow} 
\affiliation{Department of Astronomy, University of California, Berkeley, CA 94720-3411, USA}
\email{katebo@berkeley.edu}

\author[orcid=0000-0002-1092-6806, gname='Kishore',sname='Patra']{Kishore C. Patra}
\affiliation{Department of Astronomy, University of California, Berkeley, CA 94720-3411, USA}
\affiliation{Department of Astronomy \& Astrophysics, University of California, Santa Cruz, CA 95064, USA}
\email[show]{kcpatra@ucsc.edu}

\author[orcid=0009-0005-8159-8490, gname='Efrain',sname='Alvarado']{Efrain Alvarado III}
\affiliation{Department of Astronomy, University of California, Berkeley, CA 94720-3411, USA}
\affiliation{Department of Physics \& Astronomy, San Francisco State University, 1600 Holloway Avenue, San Francisco, CA 94132, USA}
\email{efrain.alvarado.iii@berkeley.edu}

\author[gname='Andreas',sname='Betz']{Andreas Betz}
\affiliation{Department of Astronomy, University of California, Berkeley, CA 94720-3411, USA}
\email{andreasbetz@berkeley.edu}

\author[gname='Victoria',sname='Brendel']{Victoria M. Brendel}
\affiliation{Department of Astronomy, University of California, Berkeley, CA 94720-3411, USA}
\email{victoriabrendel@berkeley.edu}

\author[gname='Vidhi',sname='Chander']{Vidhi Chander}
\affiliation{Department of Astronomy, University of California, Berkeley, CA 94720-3411, USA}
\email{vidhichander@berkeley.edu}

\author[gname='Asia',sname='DeGraw']{Asia A. DeGraw}
\affiliation{Department of Astronomy, University of California, Berkeley, CA 94720-3411, USA}
\email{asia.degraw@berkeley.edu}

\author[orcid=0000-0002-0498-2019, gname='Cooper',sname='Jacobus']{Cooper Jacobus}
\affiliation{Department of Astronomy, University of California, Berkeley, CA 94720-3411, USA}
\email{cjacobus@berkeley.edu}

\author[gname='Connor',sname='Jennings']{Connor F. Jennings}
\affiliation{Department of Astronomy, University of California, Berkeley, CA 94720-3411, USA}
\email{cjennings2023@berkeley.edu}

\author[gname='Ann',sname='Mina']{Ann Mina}
\affiliation{Department of Astronomy, University of California, Berkeley, CA 94720-3411, USA}
\email{annvictor2004@berkeley.edu}

\author[gname='Ansel',sname='Parke']{Ansel Parke}
\affiliation{Department of Astronomy, University of California, Berkeley, CA 94720-3411, USA}
\email{aparke@berkeley.edu}

\author[orcid=0009-0009-1267-8445, gname='Riley',sname='Patlak']{Riley Patlak}
\affiliation{Department of Astronomy, University of California, Berkeley, CA 94720-3411, USA}
\email{rileypatlak@berkeley.edu}

\author[gname='Neil',sname='Pichay']{Neil R. Pichay}
\affiliation{Department of Astronomy, University of California, Berkeley, CA 94720-3411, USA}
\email{14neil@berkeley.edu}

\author[gname='Sophia',sname='Risin']{Sophia Risin}
\affiliation{Department of Astronomy, University of California, Berkeley, CA 94720-3411, USA}
\email{sbrisin@berkeley.edu}

\author[orcid=0009-0002-2209-4813, gname='Edgar',sname='Vidal']{Edgar P. Vidal}
\affiliation{Department of Astronomy, University of California, Berkeley, CA 94720-3411, USA}
\affiliation{Department of Physics \& Astronomy, Tufts University, Medford, MA 02155, USA}
\email{e.vidal8392@berkeley.edu}

\author[gname='William',sname='Wu']{William Wu}
\affiliation{Department of Astronomy, University of California, Berkeley, CA 94720-3411, USA}
\email{wuyongxuan@berkeley.edu}

\author[orcid=0000-0001-5955-2502, gname='Thomas',sname='Brink']{Thomas G. Brink}
\affiliation{Department of Astronomy, University of California, Berkeley, CA 94720-3411, USA}
\email{tgbrink@berkeley.edu}

\author[orcid=0000-0002-2636-6508, gname='WeiKang',sname='Zheng']{WeiKang Zheng}
\affiliation{Department of Astronomy, University of California, Berkeley, CA 94720-3411, USA}
\email{weikang@berkeley.edu}

\author[orcid=0000-0003-3460-0103
, gname='Alexei',sname='Filippenko']{Alexei V. Filippenko}
\affiliation{Department of Astronomy, University of California, Berkeley, CA 94720-3411, USA}
\email{afilippenko@berkeley.edu}

\begin{abstract}

We present new transit-timing measurements for the white dwarf--cold Jupiter system WD\,1856+534, extending the baseline of observations from $311$ epochs to $1498$ epochs. The planet is unlikely to have survived the host star’s red-giant phase at its present location and is likely too small for common envelope evolution to take place. As such, a plausible explanation for the short semimajor axis is that the exoplanet started out on a much larger orbit and then spiraled inward through high-eccentricity tidal migration (HETM). A past study found tentative evidence for orbital growth, which could have been interpreted as a residual effect of HETM, but we find the data are consistent with a constant-period model after adding $18$ new transit measurements. We use the estimated period derivative $\dot{P} = 0.04\pm0.43$ ms yr$^{-1}$ to place a lower limit on the planetary tidal quality factor of $Q_p' \gtrsim 3.1 \times 10^6$, if the planet has not already achieved spin--orbit synchronization. We also test for the presence of companion planets in the system, which could have excited WD\,1856\,b onto an eccentric orbit via the Kozai--Lidov process, and ultimately rule out the presence of an additional planet with mass greater than $4.0\,M_J$ and period shorter than $1500$ days. We find no evidence for nonzero eccentricity, with an upper limit of $e\lesssim10^{-2}$. If the planet reached its current orbit through HETM, the low present-day eccentricity indicates that the migration process has now ceased, and any further orbital evolution, if any, will likely be governed by weak planetary tides.

\end{abstract}

\keywords{\uat{Exoplanets}{498} --- \uat{Exoplanet migration}{2205} --- \uat{White dwarf stars}{1799} --- \uat{Star-planet interactions}{2177} --- \uat{Transit photometry}{1709} --- \uat{Transit timing variation method}{1710}}

\section{Introduction} 
Most stars in our Galaxy host exoplanet systems and will eventually evolve into white dwarfs, yet only a handful of exoplanets orbiting white dwarfs have been detected to date \citep[see, e.g.,][]{Thorsett_1993, Luhman_2011, Gansicke_2019, Vanderburg_2020, Blackman_2021, Veras_2021, Zhang_2024}. Such systems offer a unique glimpse of what the future might hold for the nearly 6000 known planets around low-mass main-sequence stars. Many close-in planets are expected to be engulfed by the red giant phase of the host star---and indeed, an exoplanet was recently observed being consumed by its Sun-like host star \citep{De_2023}---but some may survive. At present, many questions remain unanswered about how planets that avoid engulfment will be affected by their host star's post-main-sequence evolution. For example, it is unclear how a planet could end up on an orbit so tight that it should have been consumed during the host star's red giant phase \citep{Villaver_2009, Mustill_2012, Maldonado_2020}. Characterizing the orbital properties of exoplanets around white dwarfs promises to advance our understanding of planetary migration while also helping to constrain the tidal properties of white dwarfs.

A quarter century elapsed between the discovery of the first exoplanet orbiting a main-sequence star \citep{Mayor_Queloz_1995} and the discovery of the first close-in exoplanet orbiting a white dwarf \citep{Vanderburg_2020}. In the interim, indirect evidence for the existence of planetary systems around white dwarfs continued to accumulate through observations of metal pollution in white dwarf atmospheres \citep[e.g., ][]{Zuckerman_2010, Debes_2012, Koester_2013, Koester_2014, Wilson_2014, Raddi_2015, Hollands_2017, Blouin_2019, Preval_2019} and of debris and gas disks around polluted white dwarfs \citep[e.g., ][]{Kilic_2005, Kilic_2006, Gansicke_2006, Kilic_2007, Brinkworth_2009, Brinkworth_2012, Barber_2012, Kilic_2012, Hartmann_2016, Farihi_2017}. These observations were further reinforced by the detections of a circumbinary planet orbiting the white dwarf-pulsar system PSR B1620-26 \citep{Sigurdsson_2003}, an extremely distant companion to WD 0806-661 \citep{Luhman_2011}, a disintegrating planetesimal transiting WD 1145+017 \citep{Vanderburg_2015}, and an evaporating giant planet orbiting WD J0914+1914 \citep{Gansicke_2019}. However, it was not until 2020 that \citet{Vanderburg_2020} announced the discovery of WD\,1856+534\,b, the first intact exoplanet detected on a close-in orbit around a white dwarf. 

WD\,1856+534\,b is a Jupiter-sized planet orbiting the white dwarf WD\,1856+534 (hereafter WD\,1856) at an orbital distance of merely $0.0204$\,AU \citep{Vanderburg_2020}.  Unlike most transits of close-in giant planets---which reach a transit depth of at most a few percent over an hours-long transit---the (grazing) WD\,1856\,b transit reaches a $56\%$ depth in a short $8$ minute transit. The short transit duration is typical of exoplanets around white dwarfs owing to the small radius of the star, and the large transit depth is a result of the high planet-to-star radius ratio $R_p/R_* = 7.28$ \citep{Vanderburg_2020}.

The WD\,1856 system has sparked considerable interest in the exoplanet community because the cold Jupiter lies in the host star's ``forbidden zone,'' meaning that it should not have survived the red giant phase at its present location \citep{Limbach_2025}. Many hypotheses have been proposed to explain how the planet ended up on such a tight orbit through either common envelope (CE) evolution \citep{Lagos_2020, Vanderburg_2020, Chamandy_2021, Merlov_2021, Xu_2021} or high-eccentricity tidal migration (HETM; \citealp{Munoz_2020, OConnor_2020, Vanderburg_2020, Stephan_2021}). Under CE evolution, the planet is engulfed by the expanding star's envelope and begins to spiral inward as it loses orbital energy to drag and friction within the envelope. The release of orbital energy can, in some cases, be sufficient to unbind and eject the common envelope, leaving behind an intact planet on a close-in orbit \citep{Ivanova_2020}. This mechanism is most frequently invoked to explain close binary stars, but it could also apply to the WD\,1856 system if the planet is sufficiently massive and if an additional energy source, besides the orbital energy, helps to eject the envelope \citep{Lagos_2020, Vanderburg_2020, Merlov_2021}. However, recent James Webb Space Telescope (JWST) observations targeting the WD\,1856 system have confirmed that the planet mass is at most $6\,M_J$, which is likely too small to successfully expel the envelope \citep{Limbach_2025}.

Alternatively, it is possible that WD\,1856\,b survived the red giant phase at a larger orbital distance and was brought to its present close-in orbit via tidal circularization. In order for tidal circularization to take place, the planet would first need to be excited onto a highly eccentric orbit through interactions with other bodies in or near the system. One way this could happen is through the Kozai--Lidov mechanism, in which the exchange of angular momentum between bodies in the system leads to periodic oscillations in the orbital eccentricity and inclination \citep{Kozai_1962, Lidov_1962, Naoz_2016}. Prior studies have proposed that this initial excitation could have been caused by interactions with the M dwarfs G 229-20 A/B (which form a triple-star system with WD\,1856), a stellar flyby, or additional planets in the WD\,1856 system \citep{Maldonado_2020, OConnor_2020, Vanderburg_2020, Stephan_2021}. After being excited onto a sufficiently eccentric orbit, WD\,1856\,b would have lost energy to the white dwarf owing to tidal dissipation at periapse, which would ultimately lead to a tight circular orbit as we observe today \citep{Dawson_2018, Munoz_2020, OConnor_2020}. Detecting or ruling out the presence of additional planets and residual eccentricity in the system will help us understand whether HETM is a probable explanation for WD\,1856\,b's remarkably tight orbit.

In addition to distinguishing between possible formation mechanisms, transit timing of WD\,1856\,b can also shed light on the tidal properties of the two bodies in the system. Currently, there are only two exoplanets confirmed to be undergoing orbital decay: WASP-12\,b \citep{Patra_2017, Yee_2020} and Kepler-1658\,b \citep{Vissapragada_2022}. Long-term transit-timing measurements of these two systems enabled astronomers to place new constraints on the stellar tidal quality factor, an important property describing the efficiency of tidal dissipation in the host star. The modified stellar tidal quality factor $Q_*'$ of white dwarfs remains poorly constrained because their degenerate, highly compact internal structure makes the tidal dissipation extremely small and thus difficult to measure \citep{Becker_2023}. Studies of inspiraling C-O white dwarf binaries have placed $Q_*'\approx 10^7$ \citep{Piro_2011, Burkart_2013}. A direct measurement of $Q_*'$ for an isolated white dwarf is currently lacking but is expected to lie in the range $10^{12}$--$10^{15}$ \citep{Willems_2010, Becker_2023}. If the WD\,1856 system were undergoing orbital decay or orbital growth, then we would be able to constrain the tidal properties of the white dwarf based on the observed rate of change of the period. Alternatively, if $Q_*'$ is found to be consistent with infinity, then we can interpret the evolution of the system as being driven by tides raised on the planet and can thus constrain the planetary tidal quality factor $Q_p'$ in the scenario where the planet is not yet tidally locked. Jupiter-sized exoplanets are expected to have tidal properties similar to those of our own Jupiter, but there are few tight constraints on the tidal quality factor for planets outside our solar system to date \citep[see, e.g.,][]{Efroimsky_2022, Fellay_2023, Louden_2023}.

Recent advances in direct imaging and microlensing have enabled further progress in the study of exoplanetary systems around white dwarfs. In 2023, JWST directly imaged giant planet candidates orbiting white dwarfs WD 1202-232 and WD 2105-82 \citep{Mullally_2024}. In addition, the microlensing event KMT-2020-BLG-0414 has been interpreted as a lens system consisting of a white dwarf orbited by an Earth-mass planet and a brown dwarf \citep{Zhang_2024}. Transmission spectroscopy is another area with huge potential for advancing our understanding of planetary systems around white dwarfs, particularly because of the high planet-to-star radius ratio \citep{Kaltenegger_2020}. Indeed, \citet{Xu_2021} and \citet{Alonso_2021} constrained WD\,1856\,b's mass to be greater than $0.84\,M_J$ and $2.4\,M_J$ (respectively) using transmission spectroscopy, and these constraints are likely to improve with additional JWST data. As more exoplanets are detected around white dwarfs, it becomes increasingly important to understand how these systems form and evolve through their host stars' post-main-sequence evolution. The WD\,1856 system presents a unique opportunity to do exactly that because the planet's orbit is not easily explained by any of the leading formation hypotheses.

The goal of this paper is to extend the transit-timing baseline of WD\,1856\,b in order to place tighter constraints on the system's tidal quality factor and on additional planets in the system. The remainder of the paper is organized as follows. Section~\ref{sec:methods} describes our procedures for collecting, reducing, and analyzing the new transit light curves. Section~\ref{sec:results} outlines our approach to deriving constraints on the planetary tidal quality factor and on additional planets from transit-timing variations (TTVs). We discuss our results in Section~\ref{sec:discussion} and summarize our conclusions in Section~\ref{sec:conclusions}.

\section{Methods} \label{sec:methods}
\subsection{Observations} \label{sec:observations}
We observed 20 transits of WD\,1856\,b on the 1\,m Nickel telescope at Lick Observatory atop Mount Hamilton, California. The Nickel Direct Imaging Camera is a Loral $2048\times2048$ pixel CCD with an approximately $6\overset{'}{.}3\times6\overset{'}{.}3$ field of view. Most transits were observed with $45$\,s  exposures in the Bessel $R$ filter. The filter was selected based on the white dwarf's effective temperature of $4710\,\text{ K}$ \citep[reported by][]{Vanderburg_2020}, and the exposure time was selected so as to maximize the signal-to-noise ratio while maintaining a high cadence of observations to sufficiently sample the $\sim\!8$ minute transit. The only exceptions were the 2022 April 2 transit, which was observed in the Bessel $V$ filter with $45$\,s  exposures, and the 2022 June 3 transit, which was observed in the Bessel $R$ filter with $120$\,s  exposures.

\subsection{Data Reduction}
We carried out bias subtraction and flat-field division using AstroImageJ (AIJ), an image-processing package specialized for calibrating and reducing astronomical datasets \citep{Collins_2017}. We also used AIJ to perform aperture photometry on the target star and four nearby comparison stars of similar brightness. The aperture size was chosen automatically based on the full width at half-maximum intensity: typically around $10$ pixels for the photometric aperture, $18$ pixels for the inner background annulus, and $28$ pixels for the outer background annulus. We defined the relative flux to be the ratio of the target star's flux to the sum of the comparison stars' fluxes. This relative flux was normalized to unity outside the transit. The time stamp associated with each image was taken to be the Barycentric Julian Date in Barycentric Dynamical Time ($\text{BJD}_{\text{TDB}}$) corresponding to the midexposure time.

\subsection{Light Curves} \label{sec:lightcurves}
Using the \citet{Mandel_2002} transit model, we parameterized each transit light curve by the midtransit time ($T_{\text{mid}}$), scaled stellar radius ($R_*/a$), planet-to-star radius ratio ($R_p/R_*$), and impact parameter ($b=a\cos{i}/R_*$). Though atmospheric extinction is relatively constant over the short $8$ minute transit, it can have a noticeable effect across the out-of-transit baselines on either side of the transit; as such, we introduced two additional free parameters to model the relative flux as a linear function of time. We also included a nuisance parameter $\log{f}$ to account for possible mismodeling of the uncertainties. 
 
The effect of limb darkening is challenging to model because the WD\,1856\,b transit is grazing, which means that we may not sample the full limb-darkening profile of the stellar disk over the duration of the transit \citep{Xu_2021}. Moreover, our individual transit light curves are too sparsely sampled to fit the limb-darkening coefficients as free parameters before stacking the light curves. To work around this, we took an iterative approach in which the midtransit times are first estimated by fitting the individual light curves with tabulated limb-darkening coefficients; these midtransit times are then used to stack the transit observations in order to produce a well-sampled light curve which can be fit with the limb-darkening coefficients as free parameters.

\begin{figure*}
    \centering
    \includegraphics[width=\textwidth]{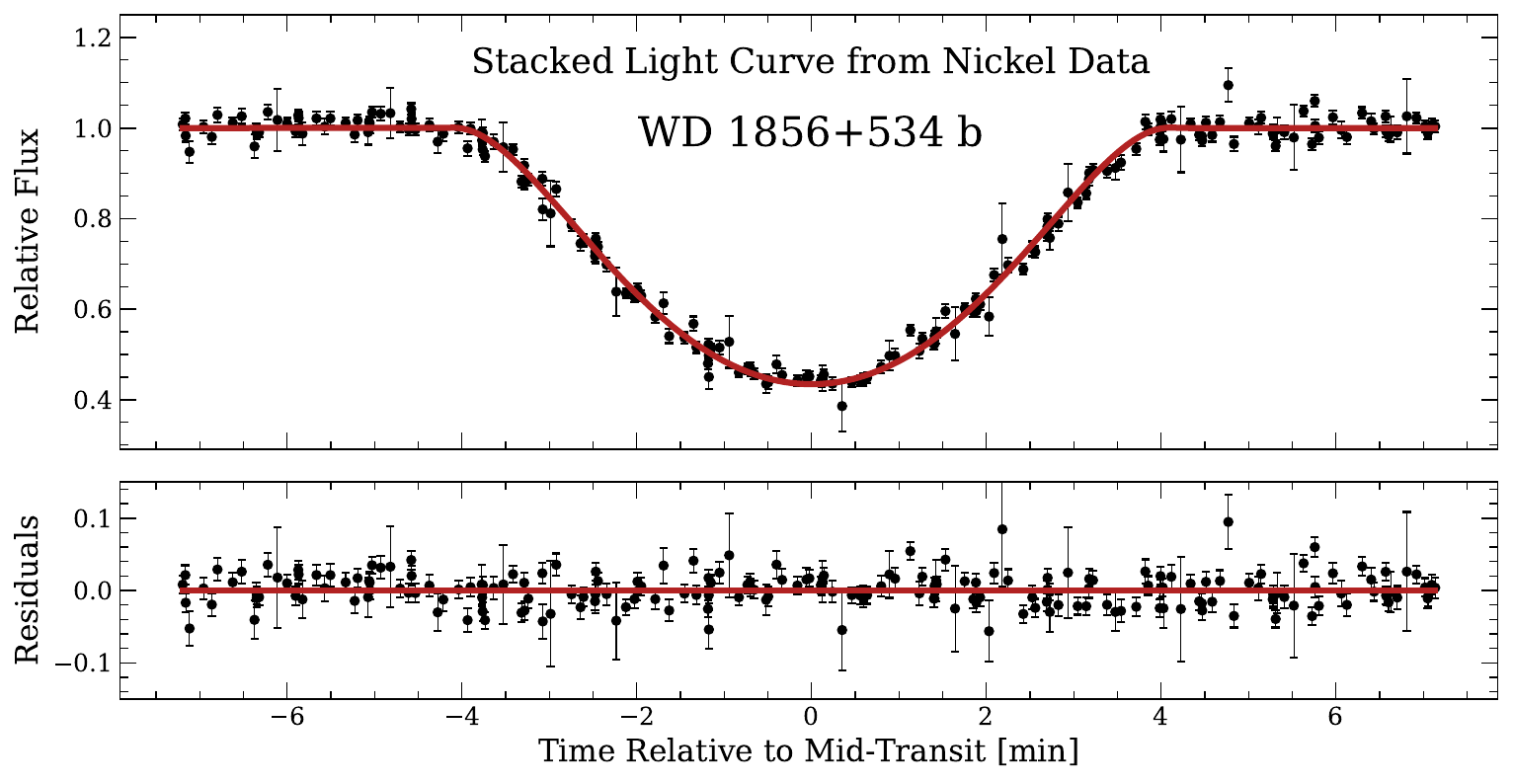}
    \caption{Stacked light curve of WD\,1856+534\,b generated from data obtained for this work.
    }
    \label{fig:stacked_lightcurve}
\end{figure*}

\begin{figure*}
    \centering
    \includegraphics[width=\textwidth]{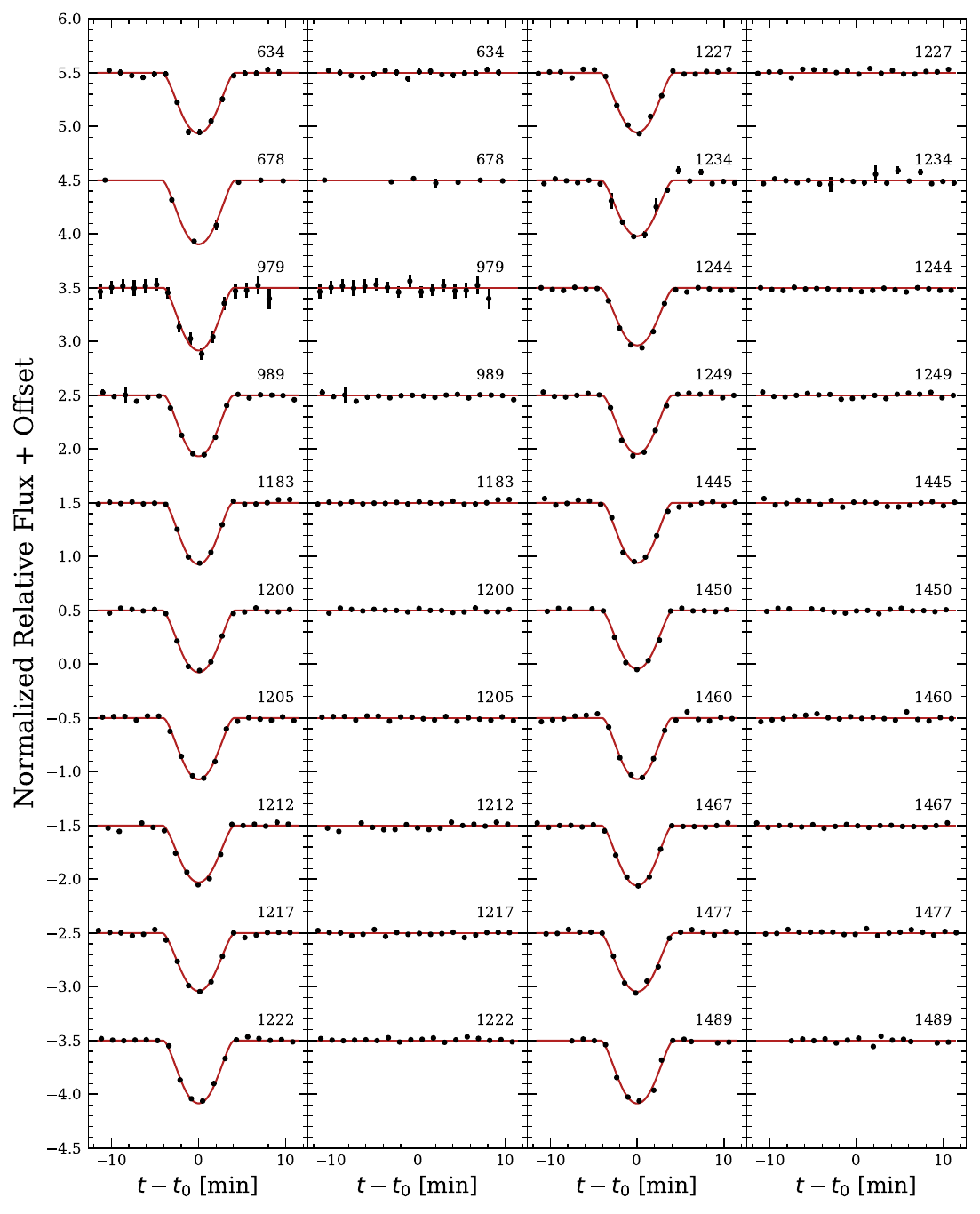}
    \caption{New WD\,1856\,b transit light curves obtained with the Lick Nickel telescope. The first and third panels show the data points (black) and fitted light curves (red) with epoch numbers labeled on the right. The second and fourth panels display the corresponding residuals. Vertical offsets were applied to separate the transits.}
    \label{fig:lightcurves}
\end{figure*}

More specifically, we performed a preliminary fit on the individual light curves assuming a quadratic limb-darkening law with the limb-darkening coefficients reported by \citet{Eastman_2013}. When calculating these coefficients, we adopted the effective temperature, $T_{\text{eff}}=4710\,\text{K}$, and surface gravity, $\log g = 7.915$, reported by \citet{Vanderburg_2020}, and the lowest metallicity included in the tables, $[\text{Fe/H}]=-4.5$. We specified uniform priors for all seven free parameters and used the \texttt{emcee} Markov Chain Monte Carlo (MCMC) ensemble sampler to approximate the posterior distributions \citep{Foreman-Mackey_2013}. Each MCMC fit was performed with $32$ walkers initialized around the maximum likelihood solution obtained from \texttt{SciPy}'s numerical optimizer \citep{SciPy_2020}. We then subtracted the optimal midtransit time from the time stamps in the individual light curves and combined the resulting data points across all transits to obtain the stacked light curve shown in Figure~\ref{fig:stacked_lightcurve}. To better constrain the physical parameters, we fit another \citet{Mandel_2002} model to the stacked light curve with the limb-darkening coefficients treated as free parameters (see Table~\ref{tab:lc_params}). Lastly, we refit the individual transit light curves with the limb-darkening parameters fixed to their optimal values as given in Table~\ref{tab:lc_params}. The resulting light curves are shown in Figure~\ref{fig:lightcurves}, and the corresponding midtransit times are reported in Table~\ref{tab:midtransit_times}. The 2022 April 2 transit was excluded from the combined light-curve fit because it was observed in a different filter; the \citet{Eastman_2013} limb-darkening parameters (rather than the optimal parameters from MCMC) were used for the final light curve of this transit. We note that these results are entirely data-driven, without any assumptions about the stellar density. This means our constraints on $b$, $R_*/a$, and $R_p/R_*$ are weaker than other results from the literature (e.g., \citealp{Vanderburg_2020, Alonso_2021, Xu_2021}), but they are not sensitive to the details of the stellar modeling. The limb-darkening parameters are also poorly constrained because the grazing transit does not fully sample the white dwarf's intensity profile.

\begin{table}
    \centering
    \caption{Median posterior values of impact parameter $b$, scaled stellar radius $R_*/a$, planet-to-star radius ratio $R_p/R_*$, and limb-darkening coefficients $u_1$ and $u_2$ derived from MCMC fit on stacked light curve. The third column lists the bounds for the uniform priors.}
    \label{tab:lc_params}
    \begin{tabularx}{0.8\linewidth}{l c c}
        \hline
        \hline
        Parameter & Value & Prior \\
        \hline
        $b$ & $8.3_{-2.7}^{+2.6}$ & $(5, 9.5)$ \\[4pt]
        $R_*/a$ & $0.00294_{-0.00036}^{+0.00058}$&$(0.002, 0.004)$\\[4pt]
        $R_p/R_*$ & $8.4_{-2.6}^{+2.5}$&$(5,12)$\\[4pt]
        $u_1$ & $0.52_{-0.32}^{+0.32}$&$(0,1)$\\[4pt]
        $u_2$ & $0.52_{-0.34}^{+0.34}$&$(0,1)$\\[4pt]
        \hline
    \end{tabularx}
\end{table}

\subsection{Orbital-change Models}\label{sec:orbital_growth}
The initial motivation for this work was to confirm or rule out secular changes in the orbital period of WD\,1856\,b. Detecting nonlinearities in the transit timing would provide further clues as to how the planet evolved onto such a tight orbit. For example, if interactions with another body in the system cause WD\,1856\,b's orbital period to vary over time, then it would be plausible that the same body was also responsible for kicking WD\,1856\,b onto its present orbit. \citet{Alvarado_2024} presented two new transit observations of WD\,1856\,b which are marginally better explained by a \textit{growing} orbit than by a constant-period model. We aim to improve upon their results by extending the transit-timing baseline over an additional $800$ orbital cycles. 

\begin{table*}
	\centering
	\caption{Best-fit values of reference transit $t_0$, orbital period $P$, and period growth rate $dP/dE$ obtained using \texttt{emcee} under constant-period and orbital-growth models. The results reported in the third and fifth columns include the TESS data, while the results reported in the second and fourth columns do not.}
	\label{tab:orbital_growth}
	\begin{tabular}{lcccc}
		\hline
        \hline
		Parameter & Constant Period & Constant Period (TESS) & Orbital Growth & Orbital Growth (TESS) \\
		\hline
            $t_0 \;\left(\text{days}\right)$ & $2459204.572725\pm10^{-6}$ & $2459204.572725\pm10^{-6}$ & $2459204.572725\pm10^{-6}$ & $2459204.572725\pm 10^{-6}$ \\
            $P\; \left(\text{days}\right)$ & $1.407939211\pm (5 \times 10^{-9})$ & $1.407939209\pm (5 \times 10^{-9})$ & $1.407939212\pm (7 \times 10^{-9}$) & $1.407939210\pm (7\times 10^{-9})$\\
            $\frac{dP}{dE}\;\left(\frac{\text{days}}{\text{epoch}}\right)$ & --- & --- & $\left(0.2\pm 1.9\right)\times 10^{-11}$ & $(0.4\pm 1.9)\times 10^{-11}$ \\
		\hline
	\end{tabular}
\end{table*}

For this analysis we combined our $20$ Nickel observations, described above, with $48$ transits recorded in the literature (see Table~\ref{tab:midtransit_times}) and $348$ transits observed by the Transiting Exoplanet Survey Satellite \citep[TESS;][]{Naponiello_2025}. The TESS data have much larger error bars than the other data points owing to the relatively long cadence of observations. For this reason, we repeated the analysis with and without the TESS data to illustrate their modest impact. We used \texttt{emcee} to fit a constant-period model and an orbital decay/growth model to the transit-timing observations \citep{Foreman-Mackey_2013}. The constant-period model is described by 
\begin{equation}
    t(E) = t_0 + P E,
\end{equation}

\noindent where $t_0$ denotes the reference transit, $P$ the period, and $E$ the epoch (orbital cycle) number. The orbital decay/growth model has the same form, but with an extra term to account for the rate of change of orbital period:
\begin{equation}
    t(E) = t_0 + P E + \frac{1}{2}\frac{dP}{dE}E^2 .
\end{equation}

Here $t_0$, $P$, and $dP/dE$ are treated as free parameters in the MCMC fits and are initialized with wide uniform priors. The best-fit values of these parameters for both the constant-period and orbital-growth models are reported in Table~\ref{tab:orbital_growth}. Based on our best-fit value of $dP/dE =\left(0.2\pm 1.9\right)\times 10^{-11}$ days/epoch, we use 
\begin{equation}
    \dot{P} = \frac{1}{P}\frac{dP}{dE}
\end{equation}

\noindent to estimate the time derivative of the period as $\dot{P} = 0.04\pm 0.43$ ms~yr$^{-1}$. Because our best-fit value of $\dot{P}$ is consistent with zero, we do not find any evidence of orbital growth or decay. However, our results can be used to place constraints on the tidal properties of the system.

\section{Results}\label{sec:results}
\subsection{Constraints on Modified Stellar Tidal Quality Factor}
\label{sec:q_star}
The modified stellar tidal quality factor $Q_*'$ describes the overall efficiency of tidal dissipation inside a star, incorporating the effects of its internal structure through the Love number \citep{Ilic_2024}. It can be shown that $Q_*'$ is related to the time derivative of the period $\dot{P}$ by
\begin{equation}
    Q_*' = \frac{27\pi}{2\vert \dot{P}\vert}\left(\frac{M_p}{M_*}\right)\left(\frac{R_*}{a}\right)^5,
\end{equation}

\noindent where $M_p$ is the mass of the planet, $M_*$ is the mass of the star, $R_*$ is the radius of the star, and $a$ is the planet's semimajor axis \citep{Goldreich_1966, Patra_2017, Alvarado_2024}. Using the most negative value of $\dot{P}$ consistent with our results in Section~\ref{sec:orbital_growth} to within $2\sigma$, the stellar mass reported by \citet{Vanderburg_2020}, the planetary mass reported by \citet{Limbach_2025}, and the scaled stellar radius derived from our MCMC fits in Section~\ref{sec:lightcurves}, we estimate that $Q_*'$ must be no smaller than $0.0034$. This is a remarkably weak limit compared to the measured $Q_*' \approx 10^{5}$ for the hot-Jupiter system WASP-12 \citep{Patra_2017, Patra_2020, Yee_2020}. The reason we are unable to constrain $Q_*'$ effectively is that the white dwarf is extremely compact, making it difficult for the planet to raise measurable tides on the star.

\subsection{Constraints on Planetary Tidal Quality Factor}
Given that we have not found evidence for ongoing orbital evolution, it is possible that the planet WD\,1856\,b  is already in spin--orbit synchronization with its host star. Indeed, most close-in giant planets are believed to be tidally locked because the timescales for spin--orbit synchronization of close-in planets are very short compared to the stellar main-sequence lifetimes \citep{Guillot_1996, Jackson_2008, Showman_2014}. If WD\,1856\,b is in a circular, aligned, spin-synchronized state, then we cannot derive useful constraints on the planetary tidal quality factor, $Q_p'$, because tides raised on the planet would produce no secular orbital evolution regardless of the value of $Q_p'$. Nonetheless, if WD\,1856\,b recently arrived onto its present close-in orbit or is being perturbed by an additional companion, then its spin may not have fully relaxed to this state. The remainder of this subsection derives a limit on $Q_p'$, assuming that the planet has not achieved spin--orbit synchronization.

The planetary tidal quality factor for a noneccentric planet in a system with no obliquity and $Q_*'=\infty$ is given by
\begin{equation}\label{eq:qp}
    Q_p' \approx \frac{27\pi}{\vert \dot{P}\vert}\left(\frac{M_*}{M_p}\right)\left(\frac{R_p}{a}\right)^5\, ,
\end{equation}

\noindent where $R_p$ denotes the planet's radius and $\dot{P}$, $M_*$, $M_p$, and $a$ are as in Section~\ref{sec:q_star} (assuming the rotation rate and mean motion are comparable but not equal; \citealp{Leconte_2010, Yang_2022}). As before, we use the most negative value of $\dot{P}$ consistent with our results in Section~\ref{sec:orbital_growth} to within $2\sigma$, the stellar mass reported by \citet{Vanderburg_2020}, the planetary mass reported by \citet{Limbach_2025}, and the scaled stellar and planetary radii derived from our MCMC fits to estimate that $Q_p'$ must be no smaller than $3.1\times 10^6$. This value is comparable to the planetary tidal quality factor of Jupiter in our own solar system, which is estimated to be between $10^5$ and $10^6$ \citep{Wu_2005}.

If the planet is not yet tidally locked but is heading toward spin--orbit synchronization, then the time for tidal spin-down can be approximated as
\begin{equation}
    \tau \approx Q_p' \left(\frac{\omega_pR_p^3}{GM_p}\right)\left(\frac{M_p}{M_*}\right)^2 \left(\frac{a}{R_p}\right)^6\, ,
\end{equation}

\noindent where $\omega_p$ denotes the planet's primordial rotation speed \citep{Guillot_1996, Showman_2014}. Adopting the minimum value of $Q_p'$ derived above and $\omega_p \approx 1.7\times 10^{-4}\, \text{ s}^{-1}$ (same as for Jupiter in our own solar system) yields a tidal synchronization timescale of at least $\tau \approx 4.4\,\text{ Myr}$. Given that planetary systems evolve on timescales of billions of years, it would be a remarkable coincidence to observe WD\,1856+534\,b within the few million years before it reached spin--orbit synchronization. Thus, the implication here is that either the planet is already tidally locked with the star or $Q_p'$ is actually much larger, $Q_p'\gtrsim10^{9}$, if the system is still tidally evolving.

\subsection{Constraints on Additional Planets from TTVs}
To place constraints on the presence of additional planets in the WD\,1856 system, we compared the observed transit times to the transit times that would be expected for a variety of hypothetical companion planets. Our analysis is inspired by that of \cite{Kubiak_2023}, who use a $431$\,day baseline of transit-timing observations to rule out the presence of a companion larger than 2\,$M_{\rm J}$ with an orbital period less than $500$\,days. Following \cite{Kubiak_2023}, we used the \texttt{TTVFast} code to efficiently simulate the expected transit times as the physical parameters of the outer planet are varied \citep{Deck_2014}. 

The orbital elements and planetary masses adopted for the \texttt{TTVFast} simulations are summarized in Table~\ref{tab:ttvfast_params}. We initialized the integrator with the stellar mass equal to the sum of the WD\,1856 mass reported by \citet{Vanderburg_2020} ($0.518\,M_{\odot}$) and the planetary mass reported by \citet{Limbach_2025} ($5.2\,M_J$). Note that it is important to include WD\,1856\,b's mass in the stellar mass because \texttt{TTVFast} assumes the planetary masses are negligible when computing the companion's orbital period, but in this system the planet-to-star mass ratio is around $1\%$ \citep{Kubiak_2023}. We set WD\,1856\,b's period and reference transit to the values obtained by fitting a linear model to our observed transit times (see Section~\ref{sec:orbital_growth}). We also assumed a circular orbit for WD\,1856\,b and adopted the inclination of $88\overset{\circ}{.}778$ reported by \citet{Vanderburg_2020}.

We began by studying hypothetical companions with masses between $\sim\!1 \, M_{\oplus}$ and $\sim\!13\,M_J$ and periods between $50$ and $1500$ days. Given that exploring a seven-dimensional parameter space with a grid search would be prohibitively expensive, we chose to focus on companions with circular orbits coplanar to that of WD\,1856\,b; that is, the companion's inclination and longitude of ascending node (LAN) are fixed to those of WD\,1856\,b. In addition, we initialized the argument of periastron (which has no physical meaning for circular orbits) to $0$ and allowed the mean anomaly of the companion to vary over all possible values.

We constructed a $300\times100\times100$ grid of evenly spaced points between the minimum and maximum values of the companion mass, period, and mean anomaly listed in Table~\ref{tab:ttvfast_params}. At each point in the grid, we computed the expected transit times for a system consisting of the star WD\,1856, the cold Jupiter WD\,1856\,b, and the hypothetical companion planet. We integrated over $2200$\,days---long enough to cover the baseline of our observations---with a $0.02$\,day time step. We then added a small correction factor to account for the Rømer delay, which is neglected in the \texttt{TTVFast} simulations but can become significant in systems with a short-period inner planet and a long-period outer planet \citep{Rappaport_2013, Kubiak_2023}. To this end, we computed the amplitude of the Rømer delay as in \citet{Rappaport_2013}:
\begin{equation}
    A_{\text{Rømer}} = \left( \frac{G}{c^{3} 4\pi^{2}}\right)^{1/3} ~ P_{\rm outer}^{2/3}~ M_{\rm outer} ~M_{\rm tot}^{-2/3} ~\sin i\, .
\end{equation}

Because we are assuming noneccentric orbits, the full time-dependent Rømer delay can be expressed as
\begin{equation}
    R(t) \approx A_{\text{Rømer}}~\sin{u(t)} = A_{\text{Rømer}}~\sin{\left[\frac{2\pi}{P_{\text{inner}}}~ (t-t_0)\right]}\, ,
\end{equation}

\noindent where we have set the eccentric anomaly \(u(t)\) equal to the mean anomaly \(M(t) = \frac{2\pi}{P_{\text{inner}}} (t-t_0)\) \citep{Rappaport_2013}. We then subtracted off the transit times predicted by the constant-period model to obtain the TTVs. From here we calculated the $\chi^2$ test statistic by comparing the observed and expected TTVs, where the uncertainties were taken to be the standard deviations on the individual midtransit times reported in Section~\ref{sec:lightcurves}.

\begin{figure*}
    \centering
    \includegraphics[width=0.95\textwidth]{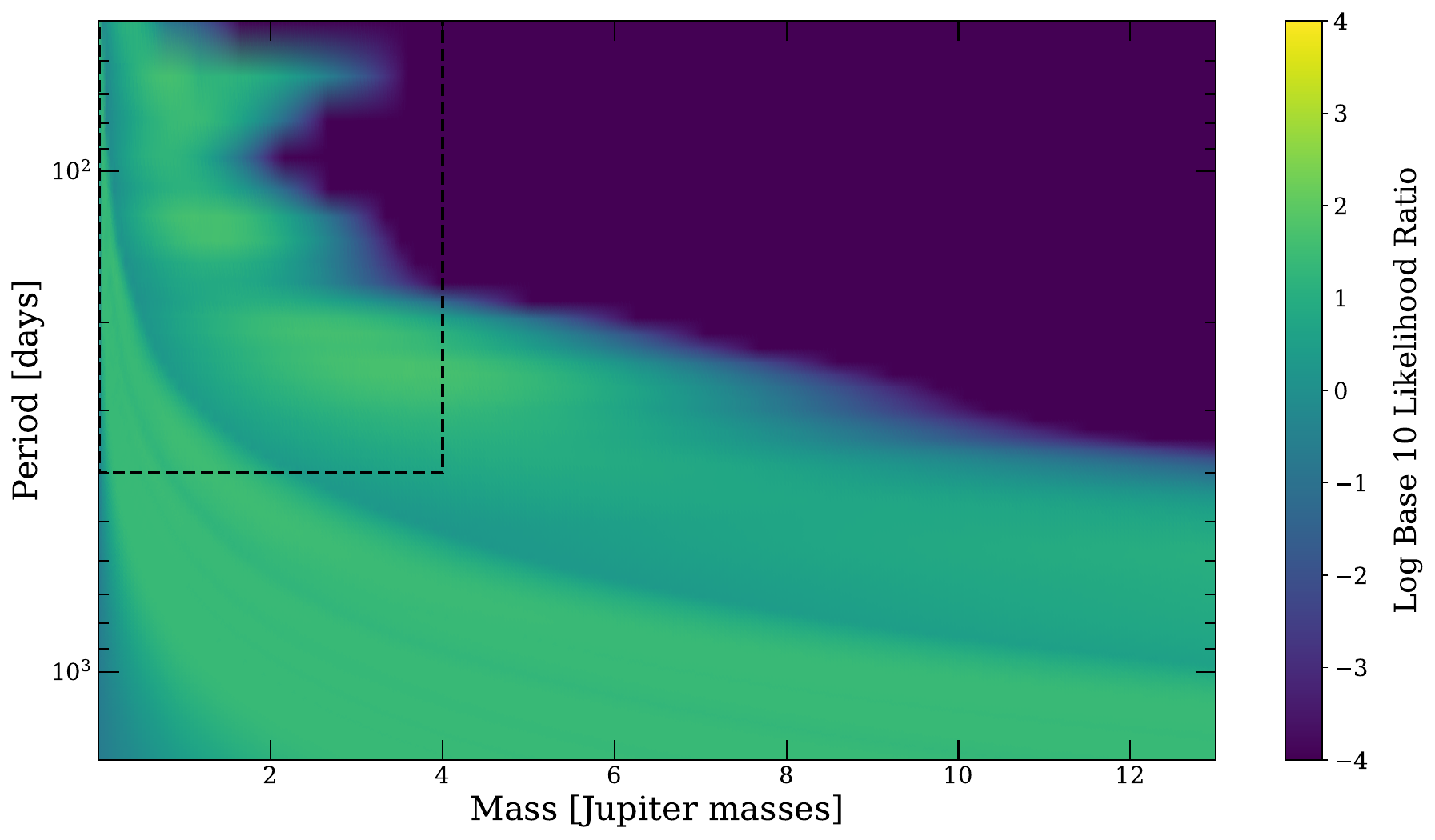}
    \caption{Log (base 10) likelihood ratio comparing hypothetical companion at $300$ mass points and $100$ period points to the constant-period model. Companion planets in the purple regions of phase space are incompatible with our data. The region outlined in a dashed box is explored further in Figure~\ref{fig:likelihood_ratios_zoom}.}
    \label{fig:likelihood_ratios}
\end{figure*}

After computing the $\chi^2$ value at each grid point, we collapsed the grid along the mean anomaly dimension by selecting the mean anomaly value that yielded the lowest $\chi^2$ for every combination of mass and period. We then defined the likelihood of each hypothetical companion as $\mathcal{L} = e^{-\chi^2/2}$ and plotted the ratio of the likelihood at each grid point to the likelihood of the constant-period model (see Figure~\ref{fig:likelihood_ratios}). Using this method, we found that the most probable hypothetical outer planet has a mass of $3.61\, M_J$ and a period of $255$\,days. The $\chi^2$ test statistic for this model is $95.68$, slightly smaller than the constant-period $\chi^2 = 103.4$.

\begin{figure*}
    \centering
    \includegraphics[width=0.95\textwidth]{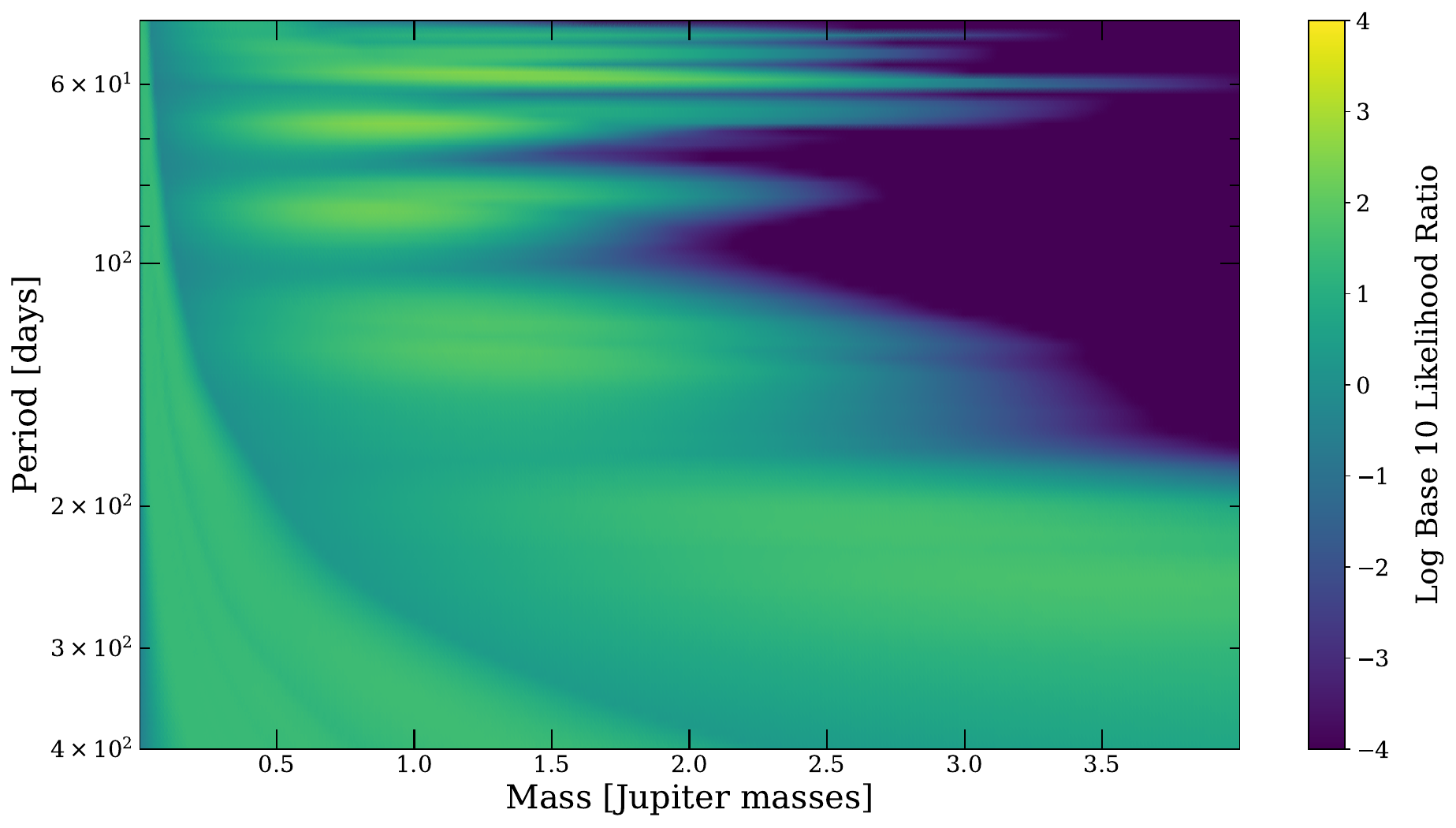}
    \caption{Log (base 10) likelihood ratio comparing hypothetical companion at $300$ mass points and $100$ period points to the constant-period model. This plot zooms in on the region outlined by a dashed box in Figure~\ref{fig:likelihood_ratios}.}
    \label{fig:likelihood_ratios_zoom}
\end{figure*}

\begin{figure*}
    \centering
    \includegraphics[width=\textwidth]{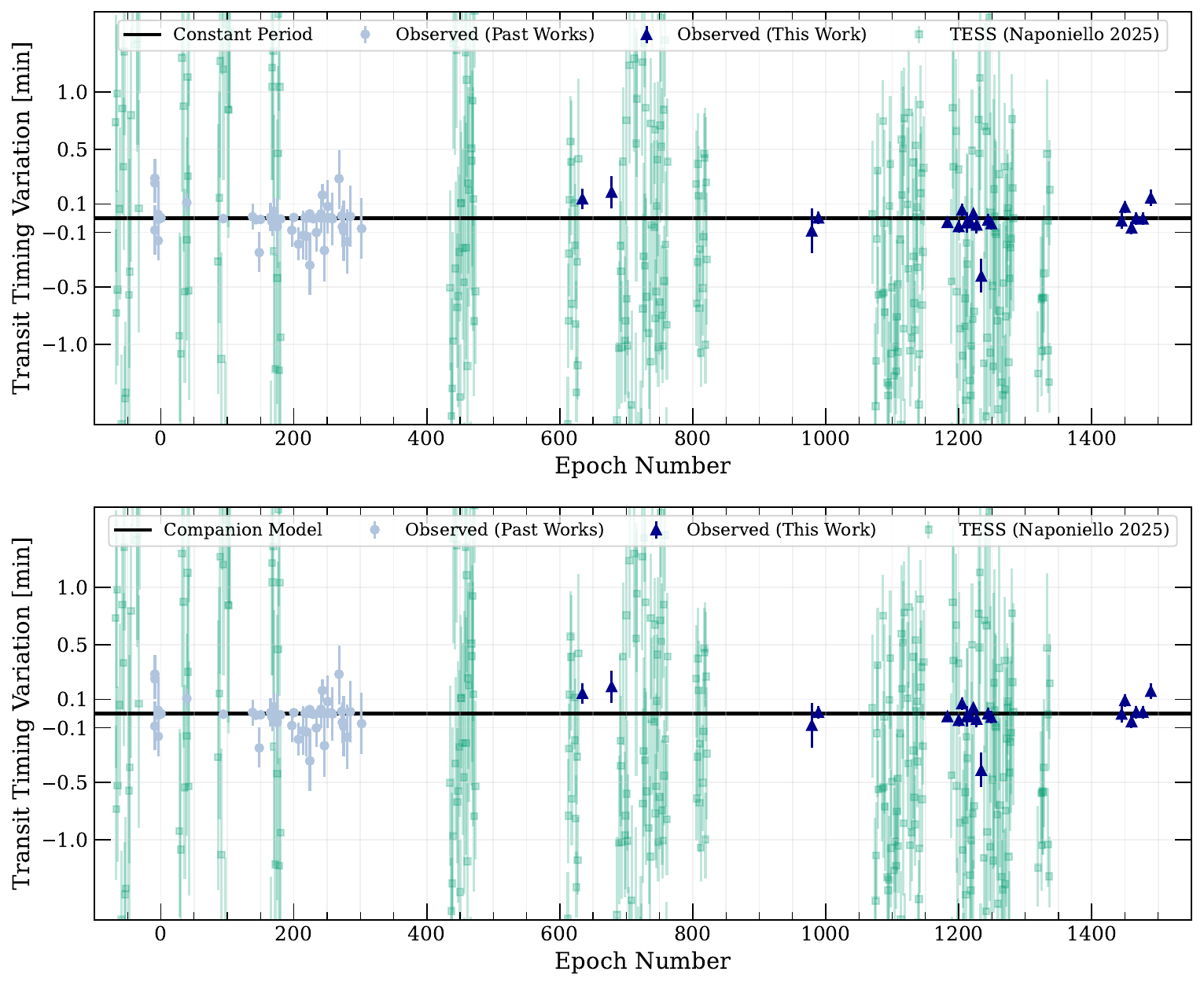}
    \caption{TTVs between observed midtransit times reported in Table~\ref{tab:midtransit_times} and expected midtransit times. In the top panel, the expected midtransit times are computed by fitting a linear (i.e., constant-period) model to the observed transit times. In the bottom panel, the expected midtransit times are obtained from the \texttt{TTVFast} simulations assuming a coplanar $0.939\,M_J$ companion on a circular orbit with a $67.1$\,day period. Note that there is no significant difference between the timing residuals, which implies that adding a companion planet to the system is not necessary to explain the observed data. The green points are midtransit times extracted from TESS data by \citet{Naponiello_2025}.}
    \label{fig:o-c}
\end{figure*}

To improve our resolution in the most probable region of phase space, we repeated the analysis on another $300\times100\times100$ grid with the maximum companion mass set to $4\,M_J$ and the maximum companion period set to $400$\,days. We also adopted a logarithmic scale for the spacing between grid points along the period dimension to more thoroughly explore the short-period region. This time the minimum $\chi^2$ value of $91.81$ occurred with a companion mass of $0.939\,M_J$ and a companion period of $67.1$\,days (see Figure~\ref{fig:likelihood_ratios_zoom}). We plotted the TTVs for this optimal companion as well as the constant-period model in Figure~\ref{fig:o-c}. Given that the improvement in the $\chi^2$ test statistic is relatively small, we do not find any evidence for additional planets in the WD\,1856 system. 

To quantify this conclusion, we computed the Bayesian information criterion (BIC) for each model evaluated in the grid search. The BIC is a statistical metric designed to compare models based on both goodness of fit and model complexity \citep{Schwarz_1978}. The BIC is defined by
\begin{equation}\label{eq:bic}
    \text{BIC} = k \ln{n} - 2 \ln{\mathcal{L}}\, ,
\end{equation}

\noindent where $k$ is the number of free parameters in the model, $n$ is the number of transit observations, and $\mathcal{L}$ is the likelihood of the model. We ultimately found that the BIC is lower for the constant-period model than for \textit{any} model including a companion planet, and that the constant-period model is preferred ($\Delta \text{BIC} > 5$) over all companion planets with masses greater than $4.0\,M_J$ and periods less than $1500$ days. 

The horizontal striations in Figure~\ref{fig:likelihood_ratios_zoom} indicate that certain periods for the companion planet are more probable than others regardless of the companion mass. One possible explanation is that these periods could correspond to orbital resonances of the system. Resonances occur either when the ratio of orbital periods for two planets is close to a ratio of small integers (mean motion resonances) or when the orbits of the two planets precess with comparable frequencies (secular resonances; \citealp{Malhotra_2012}). Horizontal striations at \textit{low} periods could be explained as mean motion resonances, but the periods of the features observed in Figure~\ref{fig:likelihood_ratios_zoom} are likely too long for this explanation. A more plausible alternative is that the observed striations are a consequence of sparsely sampled time-series data. To test this hypothesis, we repeated the \texttt{TTVFast} grid search with the TTVs shuffled so as to preserve sampling effects while scrambling any long-term trends in the transit timing. The resulting likelihood ratio plots showed some horizontal features at long periods, but did not match the striations in Figure~\ref{fig:likelihood_ratios_zoom} exactly. Nonetheless, we conclude that the striations are more likely to be explained by sampling effects than by orbital resonances.

\begin{figure}
    \centering
    \includegraphics[width=\columnwidth]{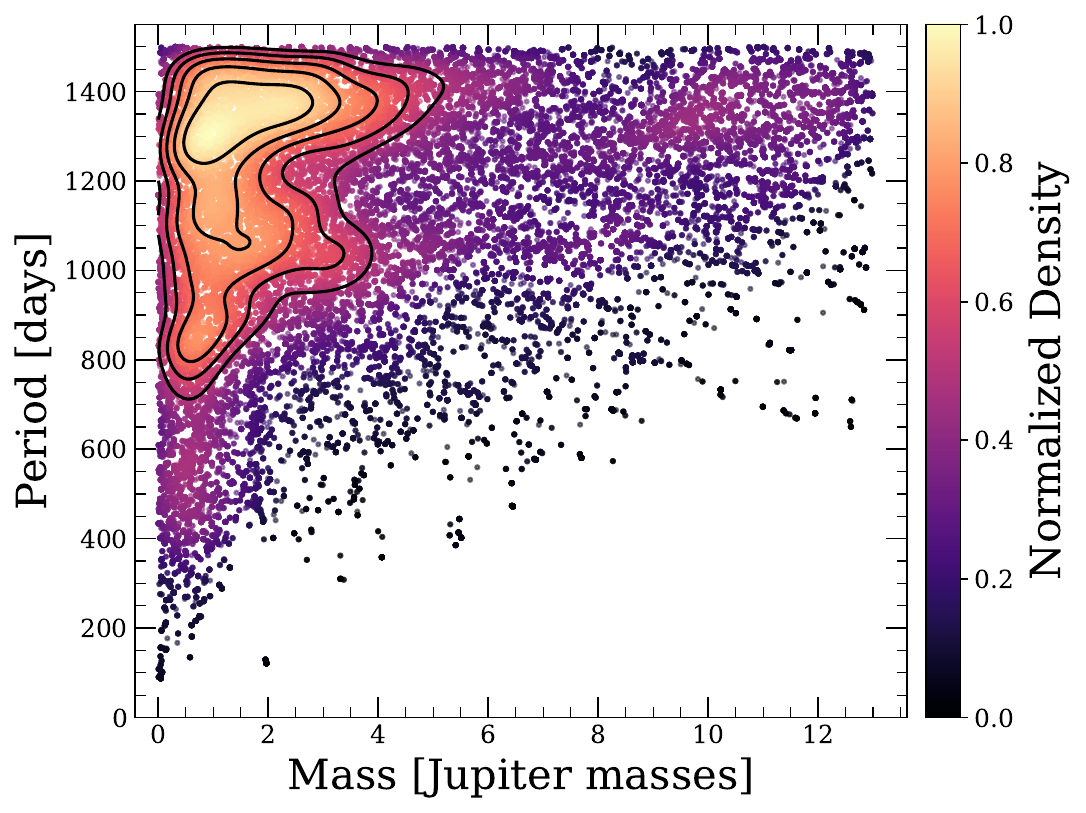}
    \caption{Two-dimensional projection of posterior distribution sampled by \texttt{emcee} when fitting the \texttt{TTVFast} transit-timing model to the observed transit times. The probability distribution function used to color points and compute contours was smoothed using Gaussian Kernel Density Estimation. The lighter colors denote regions of higher probability density. Contours are drawn at $50\%$, $60\%$, $70\%$, $80\%$, and $90\%$ of the peak probability density.}
    \label{fig:mcmc_density}
\end{figure}

Given that grid search is only feasible in low-dimensional cross sections of phase space, we have also experimented with a seven-dimensional MCMC fit built on top of the \texttt{TTVFast} simulator. The free parameters in this fit are listed in the first column of Table~\ref{tab:ttvfast_params}. We specified wide uniform priors on all seven parameters and included a nuisance parameter to account for possible mismodeling of the uncertainties. As for the light-curve and orbital-growth fits, we initialized $32$ walkers in a small Gaussian ball near the maximum likelihood solution and used \texttt{emcee} to sample the posterior distributions. The projection of the posterior distribution onto the mass and period dimensions is shown in Figure~\ref{fig:mcmc_density}. We ultimately found that the posteriors hit the prior bounds, which means we cannot infer best-fit values for the physical parameters of a hypothetical companion from our results. Moreover, the region with highest density is the low-mass, high-period region where our observations have almost no sensitivity. The lack of strong constraints on parameters other than the mass and period reinforces the grid search results by demonstrating that there are no regions of phase space with significantly higher probability that were overlooked when selecting low-dimensional cross sections for the grid search.

We chose to focus on coplanar companion planets because of limited computational resources, but it is still worthwhile to consider whether similar techniques could be applied to constrain the presence of misaligned companions. To this end, we performed several three-dimensional grid searches in alternative cross sections of phase space (i.e., the three free parameters were chosen from the first column of Table~\ref{tab:ttvfast_params} and were \textit{not} the period, mass, and mean anomaly as above). In each case, we either fixed the inclination to a value different from that of WD\,1856 or allowed the inclination to vary freely at the expense of fixing the eccentricity. These additional grid searches did not produce any more probable companion planets than the results presented earlier; nonetheless, it would be interesting to develop techniques for systematically and efficiently exploring noncoplanar companion orbits in future work.

\begin{table*}
	\centering
	\caption{Orbital elements and planetary masses used to compute expected transit times with \texttt{TTVFast}.}
	\label{tab:ttvfast_params}
	\begin{tabular}{lccc}
		\hline
            \hline
		Parameter & WD\,1856\,b & Companion Min. & Companion Max. \\
		\hline
		Mass ($M_{J}$) & $0$ & $0.003$ & $13$ \\
            Period (days) & $1.4079392$ & $50$ & $1500$ \\
            Eccentricity & 0 & 0 & 0 \\
            Inclination & $88\overset{\circ}{.}778$ & $88\overset{\circ}{.}778$ & $88\overset{\circ}{.}778$ \\
            LAN & $0^{\circ}$ & $0^{\circ}$ & $0^{\circ}$ \\
            Arg. periastron & $0$ & $0$ & $0$ \\
            Mean anomaly & $0^{\circ}$ & $0^{\circ}$ & $360^{\circ}$ \\
		\hline
	\end{tabular}
\end{table*}

\section{Discussion}\label{sec:discussion}
Although we have not found evidence for orbital evolution or additional planets in the WD\,1856 system, our results nonetheless have implications for the formation history of this system. First, we look at CE evolution. 

A CE origin for WD\,1856\,b is energetically not favored. In the standard $\alpha$-formalism, envelope ejection requires $\alpha_{\rm CE}\,\Delta E_{\rm orb} \gtrsim E_{\rm bind}$, where $E_{\rm bind}$ is the binding energy of the red giant's envelope and $\Delta E_{\rm orb}$ is the orbital energy released by the planet during inspiral \citep{Ivanova_2013}. For a typical $1\,M_\odot$ red giant progenitor with a $0.6\,M_\odot$ core, a radius of $R_\ast\approx 100\,R_\odot$, and an envelope mass $M_{\rm env}\approx 0.4\,M_\odot$, the envelope binding energy is $E_{\rm bind}\approx 3\times10^{46}\,$erg. By contrast, a $5.2\,M_{\rm J}$ planet spiraling in from large separation to the currently observed $a_f=0.02$\,AU releases only $\Delta E_{\rm orb}\approx G M_{\rm core} M_p/(2a_f)\approx 1\times10^{45}\,$erg, more than an order of magnitude too small to unbind the envelope. This would require an unphysical efficiency $\alpha_{\rm CE}\approx 25$. Thus, WD~1856\,b could not have reached its present orbit through CE evolution and must have migrated inward through a different mechanism (e.g., HETM; \citealp{Vanderburg_2020, Limbach_2025}).

A planet with a small nonzero eccentricity and tidal deformation is expected to undergo apsidal precession at a rate given by 
\begin{equation}
    \frac{d\omega}{dE} = 15\pi k_p\left(\frac{M_*}{M_p}\right)\left(\frac{R_p}{a}\right)^5
\end{equation}
\citep{Ragozzine_2009}. For the WD\,1856 system, the expected apsidal precession rate of the planet’s orbit is 
$d\omega/dE = 2.55\times10^{-5}\,\mathrm{rad\ epoch^{-1}}$, assuming a Jupiter-like Love number $k_p = 0.565$ \citep{Durante_2020}. At this precession rate, the argument of periastron advances by 
$\Delta\omega \simeq d\omega/dE\times E \simeq 2.55\times10^{-5}\times1500 \,\,\mathrm{rad}\approx 3.8\times10^{-2}\,\mathrm{rad}$ 
over the 1500 observed orbits. Hence, the current observation baseline samples a minuscule fraction of the precession cycle. 

For a slowly precessing, mildly eccentric orbit, the characteristic amplitude of the precession-induced TTV signal is of order 
$A_{\rm TTV} \approx (eP/2\pi)\,\Delta\omega$, which for $P = 1.4$ d 
($\approx 1.21\times10^{5}$ s) gives $A_{\rm TTV} \approx e \times 7.4\times10^{2}\,\mathrm{s}$. Thus, an eccentricity 
$e = 0.01$ would produce a coherent TTV signal with amplitude $\sim 7$--$8$ s across the dataset, while 
$e = 0.001$ would correspond to a subsecond signal. The combined $O-C$ diagram in Figure \ref{fig:o-c} shows no evidence for coherent curvature 
or sinusoidal structure at amplitudes $\gtrsim 5$ s over 1500 epochs. Adopting this as a conservative detection threshold 
implies $e|\cos\omega| \lesssim  7\times10^{-3}$; marginalizing 
over a uniform prior in $\omega$ then yields a robust upper limit of $e \lesssim 10^{-2}$ from the nondetection of TTVs.

This upper limit on orbital eccentricity is also consistent with an MCMC fit. An apsidally precessing orbit causes the transit times to vary according to
\begin{equation}\label{eq:apsidal_precession}
    t(E)=t_0 + P_sE - \frac{eP_a}{\pi}\cos{\omega}\, , 
\end{equation}

\noindent where $E$ is the epoch number, $t_0$ is the reference transit, $P_s$ is the sidereal period, $e$ is the eccentricity, $P_a$ is the anomalistic period, and $\omega$ is the argument of pericenter $\omega(E) = \omega_0 + (d\omega/dE)\,E$ \citep{Ragozzine_2009, Patra_2017}. The sidereal and anomalistic periods are related to each other by
\begin{equation}\label{eq:periods}
    P_s = P_a\left(1-\frac{1}{2\pi}\frac{d\omega}{dE}\right)\, . 
\end{equation}

We implemented an MCMC fit with $t_0$, $P_s$, $e$, $\omega_0$, and $d\omega/dE$ as free parameters and found that the $e=0$ scenario is strongly preferred, in which case both $\omega_0$ and $d\omega/dE$ are undefined. The low present-day eccentricity and lack of observed orbital evolution indicate that, even if HETM originally delivered the planet to its current close-in orbit, it is no longer actively operating in this system. 

Lastly, though we have focused on TTVs in this work, it is also possible to confirm or rule out the presence of companion planets through transit-\textit{duration} variations (TDVs). TDVs are fluctuations in the time from first contact to last contact caused by resonant interactions, secular precession, or stellar oblateness \citep{Agol_2018}. While TTVs are more commonly used to detect companion planets, TDVs can help to break degeneracies in the companion parameters and to study exomoons \citep{Seager_2003, Kipping_2020}. We computed the duration of each transit in our Nickel dataset as
\begin{equation}\label{eq:duration}
    T_{14} = \frac{P}{\pi}\arcsin{\left(\frac{R_*}{a}\left[\frac{\left(1+\frac{R_p}{R_*}\right)^2 - b^2}{1-\left(\frac{b R_*}{a}\right)^2}\right]^{\frac{1}{2}}\right)}
\end{equation}

\noindent and obtained values ranging from $7.9$ to $8.6$ minutes \citep{Seager_2003}. The uncertainties computed by propagating the full covariance matrices from the individual light-curve fits through Equation~\ref{eq:duration} were on the order of $\pm1$ min. In comparison, the typical TDVs produced by \texttt{TTVFast} for the closest and most massive companion planets consistent with our results are at most fractions of a second. Thus, TDVs do not provide useful constraints on the presence of additional planets in our dataset, but could be valuable for similar analyses of systems with better-sampled light curves.

\section{Conclusions}
\label{sec:conclusions}
In this work, we have analyzed transit observations of WD\,1856\,b to place constraints on the presence of orbital evolution and additional planets in the system. The WD\,1856 system offers a unique opportunity to probe planetary migration hypotheses because the planet orbits too close to its host star to have survived the host star's red giant phase at its present location. CE evolution has been proposed to explain the short semimajor axis without invoking planetary migration, but recent measurements placing the planet's mass below $6\,M_J$ make that explanation improbable. HETM offers a more natural explanation for the observed semimajor axis---but only if we posit the existence of a perturber responsible for exciting the cold Jupiter onto a highly eccentric orbit in the first place. There are many candidates for such a perturber, including companion planet(s), a stellar flyby, or the M dwarfs G 229-20 A/B, which are gravitationally bound to WD\,1856. Here we have focused on the easiest perturber to detect: an additional planet in the WD\,1856 system.

We compiled $48$ observations of WD\,1856\,b transits from the literature and $20$ new transits from our observations with the Lick Nickel telescope, $18$ of which were previously unpublished. The Nickel data extend the transit-timing baseline from $311$ to $1498$ epochs, significantly improving our ability to detect TTVs due to companion planets. We used this dataset to investigate early evidence of orbital growth reported by \citet{Alvarado_2024} and ultimately found no evidence for a changing period over a longer baseline. We also explored whether the observed TTVs are consistent with additional planets of various masses and periods using both grid search and MCMC methods. The lack of evidence for any companion planet further complicates the question of how this planetary system formed, because the most plausible formation hypothesis still requires that we invoke an unseen perturber. It is important to note that while our data allow us to rule out a large close-in planet as the perturber, we cannot yet rule out smaller close-in planets and larger planets at greater distances.

We believe tidal migration remains a plausible explanation for WD\,1856\,b's present semimajor axis, but the chances of detecting the perturber appear to be diminishing. It is not feasible to confirm or rule out the possibility of a stellar flyby, and we are unlikely to find a companion planet if it was ejected from the system in the course of the interaction that excited WD\,1856\,b onto an eccentric orbit. That being said, uncovering the formation history of the WD\,1856 system would greatly advance our understanding of planetary migration, which justifies continued transit-timing observations over an extended baseline. Our $2\sigma$ upper limit on the period derivative of $0.9\,{\rm ms \;yr}^{-1}$ indicates the planetary tidal quality factor is no smaller than $3.1\times10^6$ if the system is not yet tidally locked. Based on the lack of observed TTVs, we determined that the maximum possible eccentricity is $10^{-2}$. If HETM is responsible for bringing the cold Jupiter to its present $0.02$\,AU orbital distance, then the process has already run its course and left the planet on a nearly circularized orbit.

\begin{acknowledgments}
\indent We thank UC Berkeley undergraduate students Gracelynn Jost and Arnav Swaroop, who assisted with transit observations while training on the Nickel telescope. We are also grateful to the staff at Lick Observatory for support with the observations. Research at Lick Observatory is partially supported by a gift from Google. We thank Luca Naponiello for alerting us to the available analysis of the TESS data.

A.V.F.’s research group at UC Berkeley received financial assistance from the Christopher R. Redlich Fund, as well as donations from Gary and Cynthia Bengier, Clark and Sharon Winslow, Alan Eustace and Kathy Kwan, Timothy and Melissa Draper, Briggs and Kathleen Wood, Ellyn and Alan Seelenfreund (W.Z. is a Bengier-Winslow-Eustace Specialist in Astronomy, T.G.B. is a Draper-Wood-Seelenfreund Specialist in Astronomy),  
and numerous other donors.   
\end{acknowledgments}

\section*{Data Availability}
The raw Nickel data from 2022 to 2023 are publicly available on the Mount Hamilton Data Repository.\footnote{https://mthamilton.ucolick.org/data/} The 2024 and 2025 data will become public at the same link approximately 2\,yr  after the observation date. The analysis code is available at DOI: \href{https://doi.org/10.5281/zenodo.21267295}{10.5281/zenodo.21267295} \citep{Zenodo_Repo}.

\appendix

\section{Midtransit Times}\label{sec:transit_times}

Table~\ref{tab:midtransit_times} presents the midtransit times for the WD\,1856 system from the literature and this work.
The 20 new data points increase the total number of observations to 68 and extend the transit-timing baseline from 311 to 1498 epochs.

\begin{table*}
	\centering
    \ssmall
	\caption{Midtransit times for the
    WD\,1856 system.
    }
	\label{tab:midtransit_times}
	\begin{tabular}{lcccl}
		\hline
        \hline
		UTC Date & Epoch & $T_{\text{mid}}$ ($\text{BJD}_{\text{TDB}}$, days) & $1\sigma$ ($10^{-5}$ days) & Source \\
		\hline
		      2019 Oct 10 & $-9$ & 2458766.703572 & 12 & \citet{Vanderburg_2020}\\
            2019 Oct 10 & $-9$ & 2458766.703803 & 10 & \citet{Vanderburg_2020}\\
            2019 Oct 10 & $-9$ & 2458766.703827 & 10 & \citet{Vanderburg_2020}\\
            2019 Oct 17 & $-4$ & 2458773.743217 & 10 & \citet{Vanderburg_2020}\\
            2019 Oct 17 & $-4$ & 2458773.743321 & 19 & \citet{Vanderburg_2020}\\
            2019 Oct 17 & $-4$ & 2458773.743344 & 14 & \citet{Vanderburg_2020}\\
            2019 Oct 23 & 0 & 2458779.375085 & 0.20 & \citet{Vanderburg_2020}\\
            2019 Oct 23 & 0 & 2458779.375083 & 0.34 & \citet{Mallonn_2020}\\
            2019 Dec 17 & 39 & 2458834.284788 & 4.7 & \citet{Vanderburg_2020}\\
            2020 Mar 3 & 94 & 2458911.721367 & 0.30 & \citet{Alonso_2021}\\
            2020 May 4 & 138 & 2458973.670701 & 6.3 & \citet{Alonso_2021}\\
            2020 May 11 & 143 & 2458980.710383 & 0.80 & \citet{Alonso_2021}\\
            2020 May 18 & 148 & 2458987.749918 & 10 & \citet{Kubiak_2023}\\
            2020 May 21 & 150 & 2458990.565959 & 0.60 & \citet{Alonso_2021}\\
            2020 Jun 11 & 165 & 2459011.685057 & 6.9 & \citet{Kubiak_2023}\\
            2020 Jun 14 & 167 & 2459014.500916 & 0.70 & \citet{Alonso_2021}\\
            2020 Jun 15 & 168 & 2459015.908855 & 6.9 & \citet{Kubiak_2023}\\
            2020 Jun 15 & 168 & 2459015.908868 & 0.21 & \citet{Xu_2021}\\
            2020 Jun 15 & 168 & 2459015.908869 & 0.22 & \citet{Xu_2021}\\
            2020 Jun 18 & 170 & 2459018.724783 & 6.9 & \citet{Kubiak_2023}\\
            2020 Jun 25 & 175 & 2459025.764404 & 14 & \citet{Kubiak_2023}\\
            2020 Jul 2 & 180 & 2459032.804137 & 0.22 & \citet{Xu_2021}\\
            2020 Jul 26 & 197 & 2459056.739049 & 8.1 & \citet{Kubiak_2023}\\
            2020 Jul 30 & 200 & 2459060.962929 & 0.18 & \citet{Xu_2021}\\
            2020 Aug 9 & 207 & 2459070.818373 & 8.1 & \citet{Kubiak_2023}\\
            2020 Aug 19 & 214 & 2459080.673991 & 12 & \citet{Kubiak_2023}\\
            2020 Aug 23 & 217 & 2459084.897894 & 0.24 & \citet{Xu_2021}\\
            2020 Aug 26 & 219 & 2459087.713682 & 12 & \citet{Kubiak_2023}\\
            2020 Aug 30 & 222 & 2459091.937591 & 0.33 & \citet{Xu_2021}\\
            2020 Sep 2 & 224 & 2459094.753234 & 16 & \citet{Kubiak_2023}\\
            2020 Sep 2 & 224 & 2459094.753487 & 0.87 & \citet{Xu_2021}\\
            2020 Sep 9 & 229 & 2459101.793160 & 0.18 & \citet{Xu_2021}\\
            2020 Sep 16 & 234 & 2459108.832789 & 9.3 & \citet{Kubiak_2023}\\
            2020 Sep 23 & 239 & 2459115.872557 & 0.25 & \citet{Xu_2021}\\
            2020 Sep 26 & 241 & 2459118.688454 & 9.3 & \citet{Kubiak_2023}\\
            2020 Sep 29 & 243 & 2459121.504424 & 5.5 & \citet{Mallonn_2020}\\
            2020 Oct 3 & 246 & 2459125.727971 & 16 & \citet{Kubiak_2023}\\
            2020 Oct 3 & 246 & 2459125.728129 & 0.25 & \citet{Xu_2021}\\
            2020 Oct 10 & 251 & 2459132.767882 & 13 & \citet{Kubiak_2023}\\
            2020 Oct 17 & 256 & 2459139.807524 & 0.28 & \citet{Xu_2021}\\
            2020 Oct 20 & 258 & 2459142.623396 & 13 & \citet{Kubiak_2023}\\
            2020 Nov 3 & 268 & 2459156.702986 & 15 & \citet{Kubiak_2023}\\
            2020 Nov 9 & 272 & 2459162.334560 & 3.1 & \citet{Mallonn_2020}\\
            2020 Nov 10 & 273 & 2459163.742446 & 10 & \citet{Kubiak_2023}\\
            2020 Nov 13 & 275 & 2459166.558304 & 15 & \citet{Kubiak_2023}\\
            2020 Nov 20 & 280 & 2459173.597948 & 16 & \citet{Kubiak_2023}\\
            2020 Nov 27 & 285 & 2459180.637767 & 15 & \citet{Kubiak_2023}\\
            2020 Dec 21 & 302 & 2459204.572674 & 15 & \citet{Kubiak_2023}\\
            2022 Apr 2 & 634 & 2459672.008638 & 5.0 & This work\\
            2022 Jun 3 & 678 & 2459733.957996 & 8.0 & This work\\
            2023 Aug 1 & 979 & 2460157.747509 & 11 & This work\\
            2023 Aug 15 & 989 & 2460171.826965 & 3.0 & This work\\
            2024 May 14 & 1183 & 2460444.967150 & 2.0 & This work\\
            2024 Jun 7 & 1200 & 2460468.902096 & 3.0 & This work\\
            2024 Jun 14 & 1205 & 2460475.941873 & 3.0 & This work\\
            2024 Jun 24 & 1212 & 2460485.797385 & 5.0 & This work\\
            2024 Jul 1 & 1217 & 2460492.837087 & 3.0 & This work\\
            2024 Jul 8 & 1222 & 2460499.876821 & 2.0 & This work\\
            2024 Jul 15 & 1227 & 2460506.916462 & 4.0 & This work\\
            2024 Jul 25 & 1234 & 2460516.771781 & 9.0 & This work\\
            2024 Aug 8 & 1244 & 2460530.851453 & 3.0 & This work\\
            2024 Aug 15 & 1249 & 2460537.891132 & 3.0 & This work\\
            2025 May 18 & 1445 & 2460813.847231 & 4.0 & This work\\
            2025 May 25 & 1450 & 2460820.886993 & 3.0 & This work\\
            2025 Jun 8 & 1460 & 2460834.966284 & 3.0 & This work\\
            2025 Jun 18 & 1467 & 2460844.821905 & 3.0 & This work\\
            2025 Jul 2 & 1477 & 2460858.901295 & 3.0 & This work\\
            2025 Jul 19 & 1489 & 2460875.796668 & 4.0 & This work\\
		\hline
	\end{tabular}
\end{table*}

\bibliography{bib}{}
\bibliographystyle{aasjournalv7}

\end{document}